\documentclass[
prd,
reprint,
amsmath,amssymb,
aps,
floatfix,
]{revtex4-2}
\usepackage{graphicx}
\usepackage{amsmath}
\usepackage{amssymb}
\usepackage{natbib}
\usepackage[dvipsnames]{xcolor}     
\usepackage{ulem}
\usepackage{hyperref}
\usepackage{multirow}
\usepackage{appendix}

\newcommand{\rmthres}{{\mathrm{thres}}}
\newcommand{\rmrange}{{\mathrm{range}}}
\newcommand{\rmupper}{{\mathrm{upper}}}
\newcommand{\rmlower}{{\mathrm{lower}}}
\newcommand{\mpch}{{\mathrm{Mpc}/h}}
\newcommand{\calm}{{\mathcal{M}}}

\newcommand{\frtxt}{{$f(R)$}}

\begin{document}

\title{Using Global Gravitational Potential Weighted Correlation Function to Constrain Modified Gravity Models}

\author{Yizhao Yang}
\affiliation{State Key Laboratory of Dark Matter Physics, School of Physics and Astronomy, Shanghai Jiao Tong University, Shanghai 200240, China}
\affiliation{Key Laboratory for Particle Astrophysics and Cosmology (MOE) / Shanghai Key Laboratory for Particle Physics and Cosmology, China}

\author{Yu Yu}
\email{yuyu22@sjtu.edu.cn}
\affiliation{State Key Laboratory of Dark Matter Physics, School of Physics and Astronomy, Shanghai Jiao Tong University, Shanghai 200240, China}
\affiliation{Key Laboratory for Particle Astrophysics and Cosmology (MOE) / Shanghai Key Laboratory for Particle Physics and Cosmology, China}

\author{Pengjie Zhang}
\affiliation{State Key Laboratory of Dark Matter Physics, School of Physics and Astronomy, Shanghai Jiao Tong University, Shanghai 200240, China}
\affiliation{Tsung-Dao Lee Institute, Shanghai Jiao Tong University, Shanghai 200240, China}
\affiliation{Key Laboratory for Particle Astrophysics and Cosmology (MOE) / Shanghai Key Laboratory for Particle Physics and Cosmology, China}

\date{\today}

\label{firstpage}


\begin{abstract}
We propose a new marked two-point correlation function weighted by the global gravitational potential as a probe for testing gravity models. Using the $\Lambda$CDM model based on general relativity (GR) as a reference, we investigate two representative modified gravity (MG) scenarios: $f(R)$ gravity and nDGP. The mark used in this work, the global gravitational potential that is reconstructed from the galaxy distribution via the Poisson equation, is in contrast to the local property based mark (e.g., local galaxy number density or gravitational potential of host halo) used in previous studies.
By applying two weighting schemes to quantify environment-dependent clustering, we find that this statistic is able to distinguish MG models from GR, with the signal being enhanced in regions corresponding to particular ranges of gravitational potential. These results indicate that the proposed statistic can serve as a useful complement to conventional clustering probes in future surveys, once observational effects and modeling uncertainties are properly taken into account.
\end{abstract}


\maketitle

\section{Introduction}\label{introduction}

The discovery of cosmic acceleration from Type Ia supernovae observations
\citep{riess_observational_1998,perlmutter_measurements_1999}
implies that a cosmological model based on General Relativity (GR) with only the known matter and radiation components is incomplete.
Within the standard GR framework, the simplest remedy is to introduce an additional smooth component with negative pressure, commonly parameterized by a cosmological constant $\Lambda$ (i.e., $\Lambda$CDM model), or more generally by dynamical dark energy
\citep{weinberg_observational_2013,olive_review_2014}.
An alternative possibility is that gravity itself is modified on cosmological scales, while remaining consistent with the stringent tests of GR in the Solar System and other high-density environments; see
Refs.~\citep{clifton_modified_2012,joyce_beyond_2015,koyama_cosmological_2016,ishak_testing_2019,hou_cosmological_2023}
for reviews of modified-gravity (MG) scenarios.

Confirming or refuting these two categories of explanations is one of the main goals of the ongoing or forth-coming Stage-IV galaxy redshift surveys including
DESI \citep{desi_collaboration_aghamousa_desi_2016},
Euclid \citep{laureijs_euclid_2011},
LSST \citep{lsst_dark_energy_science_collaboration_mandelbaum_lsst_2021},
WFIRST \citep{spergel_wide-field_2015}, 
and CSST \citep{gong_future_2025}.
These surveys will map the galaxy 3D distribution with unprecedented precision in extremely
large volumes.
The recent detection of gravitational waves
\citep{ligo_scientific_collaboration_gravitational_2017}
from the binary neutron star merger GW170817, coupled with the simultaneous measurement of its optical counterpart GRB170817A, has several popular classes of model been ruled out. However, a variety of other theories remain viable. Therefore, it remains important to test or constrain these gravity models at cosmological scales.

In this paper, we examine two representative instances of modified gravity models: chameleon $f(R)$ gravity
\citep{carroll_is_2004,carroll_cosmology_2005}
and the 5D brane-world Dvali-Gabadadze-Porrati (DGP) model
\citep{dvali_4d_2000}, specifically focusing on the normal branch (nDGP). 
We have selected these models as they offer an ideal framework for testing gravity on cosmological scales for several reasons: (i) they embody the potential to enable screening mechanisms to pass local gravity tests, whilst leaving discernible markers on broader cosmological scales. Each model encapsulates a distinct screening mechanism, namely, the chameleon
\citep{khoury_chameleon_2004}
and Vainshtein
\citep{vainshtein_problem_1972}
mechanism, (ii) the cosmological performance of these models, with differing parameters, qualitatively is representative of that of other model classes, (iii) these models have been comprehensively analyzed to date, and the essential simulations required for their testing using galaxy surveys have been developed and refined.

The screening mechanism is a common feature of the surviving MG models. The fifth force introduced by these models is effectively suppressed in high-density regions (in the case of $f(R)$ gravity) or within the Vainshtein radius (in the case of the nDGP model). This suppression allows MG models to conform with local or Solar-scale gravity tests. However, in under-dense regions or outside the Vainshtein radius, the unscreened fifth force alters structure formation, embodying the essence of the screening mechanism which is detectable. This unique mechanism presents opportunities to test these MG models utilizing statistical methods that can extract information from our regions of interest. In this study, we explore the marked correlation method to test gravity, using the $f(R)$ and nDGP models as examples.

The marked correlation function (MCF) \citep{beisbart_luminosity-_2000,beisbart_mark_2002} is a higher-order statistic that extends the standard two-point galaxy correlation function by incorporating information about galaxy properties. In the MCF framework, each galaxy is assigned a weight (the ``mark'') derived from one or more observables (e.g., local environment, luminosity, or other tracer properties). Correlating these marks as a function of separation highlights property-dependent clustering, thereby enhancing the sensitivity of clustering analyses to subtle, environmentally driven effects.
This makes the MCF a promising probe for distinguishing MG scenarios from $\Lambda$CDM. In this work, we analyze galaxy cataloges from $f(R)$ and nDGP simulations alongside matched $\Lambda$CDM cataloges that are fine tuned to reproduce similar two-point clustering. We then investigate whether suitably chosen marks enable the MCF to discriminate between these models.


MCF have been recently used in several studies to constrain gravity models (see
Ref.~\cite{alam_towards_2021} for review),
based on perturbation theory
\citep{white_marked_2016,aviles_marked_2020} 
and emulator
\citep{armijo_testing_2018,hernandez-aguayo_marked_2018}. 
These studies primarily base their mark definitions on local density and properties of the host halo. However, the screening mechanism in $f(R)$ gravity primarily depends on the depth of the potential well
\citep{he_revisiting_2014}. 
It's noteworthy that the gravitational potential at any given position comprises both the potential attributed to the host galaxy halo, which can be estimated analytically assuming the Navarro–Frenk–White (NFW) density profile, and the potential induced by matter outside the host halo, dubbed as `experienced gravity' in Ref.~\cite{shi_environmental_2017}. 
Our primary objective is to evaluate the ability of MCF to differ gravity models, wherein the mark is defined based on the global gravitational potential.

The outline of this paper is as follows: Section \ref{Modified Gravity Models} provides an introduction to the $f(R)$ gravity and nDGP models. Section \ref{Simulation and Catalogs} details the simulations and galaxy catalogues used in this study. Section \ref{MCF cal and result} presents the computation and results of the marked correlation function. Finally, we conclude the paper in section \ref{Conclusion}.

\section{Modified Gravity Models}\label{Modified Gravity Models}
MG models aim to explain the late-time acceleration of the universe by modifying the gravity itself. The $f(R)$ gravity accomplishes this by adding an additional function of the Ricci curvature, $f(R)$, to the standard Einstein-Hilbert action. Meanwhile, the nDGP models assume that gravity live within a 3+1 dimensional space, with the additional dimension having a considerably extensive length scale $r_c$.

\subsection{\frtxt{} Gravity}

We start from the modified Einstein--Hilbert action that defines \frtxt{} gravity,
\begin{equation}
S=\int\mathrm{d}^4x\sqrt{-g}\left[\frac{1}{2\kappa^2}\bigl(R+f(R)\bigr)+\mathcal{L}_m\right],
\label{f(R) action}
\end{equation}
where $g$ is the determinant of the metric tensor $g_{\mu\nu}$, $\kappa^2\equiv8\pi G$, $\mathcal{L}_m$ is the Lagrangian density of matter, and $f(R)$ is a nonlinear function of the Ricci scalar $R$.

Varying Eq.~(\ref{f(R) action}) with respect to $g_{\mu\nu}$ yields the modified Einstein equations. Since the Lagrangian depends nonlinearly on $R$, the variation naturally introduces
\begin{equation}
f_R \equiv \frac{\mathrm{d}f(R)}{\mathrm{d}R},
\end{equation}
together with its derivatives. As a result, $f_R$ is not merely a shorthand derivative but an additional scalar degree of freedom in metric $f(R)$ gravity, commonly called the \textit{scalaron}. This becomes explicit by taking the trace of the modified Einstein equations, which gives a Klein--Gordon--type equation for $f_R$,
\begin{equation}
3\,\Box f_R = R - f_R R + 2f(R) - 8\pi G\rho_m,
\label{trace_fR}
\end{equation}
where $\Box\equiv\nabla^\mu\nabla_\mu$ is the covariant d'Alembertian. Equivalently, one may define an effective potential $V_{\rm eff}(f_R)$ through
\begin{equation}
\frac{\mathrm{d}V_{\rm eff}}{\mathrm{d}f_R}
= -\frac{1}{3}\left(R - f_R R + 2f(R) - 8\pi G\rho_m\right),
\label{Veff_def}
\end{equation}
which makes clear that the local matter density shifts the minimum of $V_{\rm eff}$ and hence changes the scalaron effective mass.

In the quasi-static and weak-field limits \citep{bose_testing_2015}, the modified Poisson equation can be written as
\begin{equation}
\nabla^2\Psi=4\pi Ga^2\delta\rho_m-\frac{1}{2}\nabla^2f_R,
\label{f(R) poisson}
\end{equation}
where $\Psi$ is the Newtonian potential and $\delta\rho_m\equiv\rho_m-\bar{\rho}_m$ is the matter density perturbation. The extra term $-\tfrac12\nabla^2 f_R$ represents the scalaron contribution to the dynamical potential, i.e.\ the (screenable) fifth-force component mediated by $f_R$.

Applying the same limits to the trace equation (Eq.~\ref{trace_fR}) and subtracting the background, $\Box f_R\simeq +\nabla^2 f_R/a^2$ leads to
\begin{equation}
\nabla^2f_R=+\frac{a^2}{3}\delta R-\frac{8\pi Ga^2}{3}\delta\rho_m,
\label{f_R poisson}
\end{equation}
where $\delta R\equiv R-\bar{R}$ is the Ricci-scalar perturbation. Physically, $\delta\rho_m$ acts as an external source for the scalaron, while $\delta R$ encodes the nonlinear self-interaction through the implicit relation $R=R(f_R)$ (or equivalently $f_R=f_R(R)$). This nonlinearity is essential for the screening behavior discussed below.

\subsubsection{The Chameleon mechanism}

Equation~(\ref{f(R) action}) introduces the scalaron $f_R$ into the gravitational sector, which generically gives rise to an additional ``fifth force''. To satisfy local tests of gravity, $f(R)$ models typically rely on the chameleon screening mechanism \citep{khoury_chameleon_2004}, in which the scalaron becomes heavy in dense environments and the fifth force becomes short-ranged.

Combining Eq.~(\ref{f(R) poisson}) and Eq.~(\ref{f_R poisson}) eliminates $\nabla^2 f_R$ and gives
\begin{equation}
\nabla^2\Psi=\frac{16\pi Ga^2}{3}\delta\rho_m+\frac{a^2}{6}\delta R.
\label{combination}
\end{equation}
Two limiting cases of Eq.~(\ref{combination}) illustrate the screening behavior:
\begin{itemize}
\item \textbf{High-density (screened) regions:} the scalaron sits close to the minimum of $V_{\rm eff}$ and becomes heavy, driving the solution towards
$\delta R\simeq -8\pi G\,\delta\rho_m$.
Eq.~(\ref{combination}) then reduces to the standard Poisson equation,
\begin{equation}
\nabla^2\Psi=4\pi Ga^2\delta\rho_m,
\end{equation}
indicating that the fifth force is efficiently suppressed.

\item \textbf{Low-density (unscreened) regions:} when the local density is low, the scalaron is light and its Compton wavelength $\lambda_C \equiv m_{f_R}^{-1}$ greatly exceeds the scales of interest. In this limit the mass term $\delta R(f_R)$ in Eq.~(\ref{f_R poisson}) becomes negligible compared with the source term $\delta\rho_m$, so that $\delta R \approx 0$ is a good approximation. Eq.~(\ref{combination}) then becomes
\begin{equation}
\nabla^2\Psi=\frac{16}{3}\pi Ga^2\delta\rho_m,
\end{equation}
corresponding to an enhancement of the effective gravitational coupling by a factor $4/3$ relative to GR, which can modify structure formation.

\end{itemize}

\subsubsection{The Hu \& Sawicki model}

For concreteness, we adopt the Hu \& Sawicki form \citep{hu_models_2007},
\begin{equation}
f(R)=-m^2\frac{c_1(R/m^2)^n}{c_2(R/m^2)^n+1},
\end{equation}
where the mass scale is $m^2=\kappa^2\overline{\rho}_0/3$, with $\overline{\rho}_0$ the mean matter density today, and $c_1$ and $c_2$ are dimensionless parameters.

To produce an expansion history close to $\Lambda$CDM, one typically requires $|f_{R0}|\ll 1$, which corresponds to the high-curvature regime $R_0\gg m^2$. In this limit,
\begin{equation}
\lim_{m^2/R \to 0}f(R)\approx-\frac{c_1}{c_2}m^2+\frac{c_1}{c_2^2}m^2\left(\frac{m^2}{R}\right)^n,
\end{equation}
and hence
\begin{equation}
f_R\approx-n\frac{c_1}{c_2^2}\left(\frac{m^2}{R}\right)^{n+1}.
\label{7}
\end{equation}

In the chameleon regime the scalaron stays close to the minimum of $V_{\rm eff}$, so $R$ approximately tracks the local matter density. To leading order one may write
\begin{equation}
R \simeq \kappa^2\rho_m - 2f(R)
\approx \kappa^2\rho_m + 2\frac{c_1}{c_2}m^2,
\end{equation}
and matching the $\Lambda$CDM background expansion gives
\begin{equation}
\frac{c_1}{c_2}\approx 6\frac{\Omega_\Lambda}{\Omega_m}.
\end{equation}
The model is then conveniently specified by $(n,f_{R0})$ rather than $(n,c_1/c_2^2)$. Using Eq.~(\ref{7}) evaluated at $z=0$ and the $\Lambda$CDM-like relation $\bar{R}_0 \simeq 3m^2\left(1+4\frac{\Omega_\Lambda}{\Omega_m}\right)$, one obtains
\begin{equation}
\frac{c_1}{c_2^2}=-\frac{1}{n}\left[3\left(1+4\frac{\Omega_\Lambda}{\Omega_m}\right)\right]^{n+1}f_{R0}.
\end{equation}
In this paper we fix $n=1$ following Ref.~\citep{alam_towards_2021}.

\subsection{DGP model}
We now briefly introduce the Dvali-Gabadadze-Porrati (DGP) braneworld model. In this scenario our observable Universe is a four-dimensional hypersurface (the ``brane'') embedded in a five-dimensional bulk spacetime. Matter fields are confined to the brane, while gravity propagates in the full 5D bulk. A distinctive ingredient of DGP is an induced 4D Einstein-Hilbert term on the brane, so that gravity behaves effectively 4D on sufficiently small scales, but gradually ``leaks'' into the extra dimension on very large scales.

The DGP action can be written as
\begin{equation}
\begin{split}
S &=
\frac{1}{2\kappa^2}\int_{\rm bulk}\! \mathrm{d}^5x\,\sqrt{-g^{(5)}}\,R^{(5)}
+ \frac{1}{2\kappa_4^2}\int_{\rm brane}\! \mathrm{d}^4x\,\sqrt{-\gamma}\,R^{(4)} \\
&+ \int_{\rm brane}\! \mathrm{d}^4x\,\sqrt{-\gamma}\,\mathcal{L}_m 
+ \frac{1}{\kappa^2}\int_{\rm brane}\! \mathrm{d}^4x\,\sqrt{-\gamma}\,K \, .
\end{split}
\end{equation}
where $g^{(5)}_{AB}$ is the 5D metric, $\gamma_{\mu\nu}$ is the induced 4D metric on the brane, and $K_{\mu\nu}$ is the extrinsic curvature with trace $K\equiv K^\mu{}_\mu$. The last term is the (Gibbons--Hawking--York) boundary term that ensures a well-posed variational principle. We use $\kappa_4^2\equiv 8\pi G$.

The transition between 4D and 5D gravity is governed by the crossover scale
\begin{equation}
r_c \equiv \frac{\kappa_5^2}{2\kappa_4^2}.
\end{equation}
For distances $r\ll r_c$ gravity is effectively four-dimensional on the brane, whereas for $r\gg r_c$ gravitational dynamics becomes sensitive to the extra dimension.

Solving the 5D Einstein equations yields the standard continuity equation for brane matter,
\begin{equation}
\dot{\rho}+3H(\rho+p)=0,
\end{equation}
but a modified Friedmann equation with two branches. In the spatially flat case (which is typically adopted in large-scale structure simulations), the modified Friedmann equation can be expressed schematically as
\begin{equation}
H^2 \pm \frac{H}{r_c} = \frac{8\pi G}{3}\rho_{\rm tot},
\end{equation}
where $+$($-$) corresponds to the self-accelerating (normal) branch. The self-accelerating branch admits a late-time de Sitter solution even without a cosmological constant, $H\rightarrow 1/r_c$, but it is plagued by a ghost instability in the helicity-0 sector. For this reason we focus on the normal-branch solution, known as nDGP. Since the normal branch does not self-accelerate by itself, one usually introduces a smooth dark-energy component (or $\Lambda$) and, if desired, tunes its equation of state so that the background expansion matches $\Lambda$CDM; in that case, the deviations from GR arise predominantly from the modified growth of perturbations rather than from a different $H(a)$.

On sub-horizon scales and in the quasi-static, weak-field regime, scalar perturbations can be described by the Newtonian potential $\Psi$ together with an additional scalar degree of freedom $\varphi$, the brane-bending mode (equivalently, the helicity-0 polarization of the 5D graviton). In this limit the potential on the brane can be written as
\begin{equation}
\nabla^2\Psi = \nabla^2\Psi_N + \frac{1}{2}\nabla^2\varphi
            = 4\pi G a^2 \delta\rho_m + \frac{1}{2}\nabla^2\varphi,
\end{equation}
where $\Psi_N$ is the usual Newtonian potential in GR. The scalar field obeys a non-linear equation of motion,
\begin{equation}
\nabla^2\varphi
+\frac{r_c^2}{3\beta(a)a^2}\left[(\nabla^2\varphi)^2-(\nabla_i\nabla_j\varphi)^2\right]
=\frac{8\pi G a^2}{3\beta(a)}\,\delta\rho_m,
\label{eq:nDGP_scalar}
\end{equation}
with the time-dependent coupling function
\begin{equation}
\beta(a)=1+2Hr_c\left(1+\frac{\dot{H}}{3H^2}\right).
\end{equation}
The structure of Eq.~\eqref{eq:nDGP_scalar} is crucial: the extra force is mediated by $\varphi$, but it contains derivative self-interactions (the term quadratic in second derivatives) that become important in non-linear environments and restore GR via the Vainshtein mechanism (see below).

If we keep only the leading (linear) contribution of $\varphi$, Eq.~\eqref{eq:nDGP_scalar} reduces to $\nabla^2\varphi \simeq (8\pi G a^2/3\beta)\delta\rho_m$, and therefore
\begin{equation}
\nabla^2\Psi \simeq 4\pi G a^2 \left(1+\frac{1}{3\beta(a)}\right)\delta\rho_m.
\end{equation}
Thus, in the linear regime the effective gravitational coupling for the growth of structure is enhanced by a factor $1+1/(3\beta)$ (for nDGP, $\beta>0$), while on small scales the enhancement must be screened to satisfy local gravity constraints.

\subsubsection{The Vainshtein mechanism}
The Vainshtein mechanism in DGP is a derivative screening effect: near massive objects (or in sufficiently non-linear regions) the non-linear derivative interactions in Eq.~\eqref{eq:nDGP_scalar} dominate over the linear term. This suppresses spatial gradients of $\varphi$ and hence suppresses the ``fifth force'' contribution $\propto \nabla(\varphi/2)$, allowing the metric potentials to approach their GR values.

A useful analytic illustration is provided by a spherically symmetric mass distribution. In this case Eq.~\eqref{eq:nDGP_scalar} can be integrated to give
\begin{equation}
\frac{\mathrm{d}\varphi}{\mathrm{d}r}
=\frac{Gm(r)}{r^2}\,\frac{4}{3\beta}\,g\!\left(\frac{r}{r_*}\right),
\qquad
g(x)=x^3\left(\sqrt{1+x^{-3}}-1\right),
\end{equation}
where $m(r)$ is the enclosed mass fluctuation and $r_*$ is defined by
\begin{equation}
r_*(r)=\left(\frac{16Gm(r)\,r_c^2}{9\beta^2}\right)^{1/3}.
\end{equation}
Outside the matter distribution $m(r)=\mathrm{const.}$, the scale $r_*$ becomes a constant and is identified as the Vainshtein radius. The ratio between the fifth force and the Newtonian force is then
\begin{equation}
\frac{F_{\rm 5th}}{F_N}
=\frac{1}{2}\frac{\mathrm{d}\varphi/\mathrm{d}r}{\mathrm{d}\Psi_N/\mathrm{d}r}
=\frac{2}{3\beta}\,g\!\left(\frac{r}{r_*}\right).
\end{equation}
This makes the screening behavior transparent:
(i) for $r\gg r_*$, $g(x)\rightarrow 1/2$ and $F_{\rm 5th}/F_N\rightarrow 1/(3\beta)$, i.e.\ the linear modification is recovered;
(ii) for $r\ll r_*$, $g(x)\sim x^{3/2}$ so $F_{\rm 5th}/F_N \propto (r/r_*)^{3/2}\!\rightarrow 0$, i.e.\ the fifth force is strongly suppressed and GR is effectively restored.

Beyond spherical symmetry, Vainshtein screening is controlled by the non-linear operator
$(\nabla^2\varphi)^2-(\nabla_i\nabla_j\varphi)^2$ in Eq.~(\ref{eq:nDGP_scalar}).
Since it depends on second derivatives of $\varphi$, sufficiently long-wavelength external fields can be locally approximated as nearly constant gradients and hence are only weakly screened, allowing large-scale dynamics to remain close to linear theory while small-scale fifth forces are suppressed.

\section{Simulations and Mock Galaxy Catalogs}\label{Simulation and Catalogs}

We use the ELEPHANT (Extended LEnsing PHysics using ANalaytic ray Tracing) simulation suite, which was produced with the ECOSMOG code \citep{li_ecosmog_2012,bose_speeding_2017} for $f(R)$ gravity and its extension ECOSMOG-V \citep{li_exploring_2013,barreira_speeding_2015} for nDGP gravity. 
All simulations were run in a cubic box of side length $1024\,h^{-1}\mathrm{Mpc}$ with $1024^3$ dark matter particles of mass $m_p \approx 7.8 \times 10^{10}\,h^{-1}M_\odot$. 
Across all gravity models, the simulations adopt the best-fitting cosmological parameters from the WMAP 9-year results \citep{hinshaw_nine-year_2013}. 

The ELEPHANT suite provides five independent realizations, each corresponding to a distinct set of initial conditions (i.e., a different random phase realization of the initial density field). 
Within each realization, simulations are available for GR as well as for a set of modified-gravity models. 
For $f(R)$ gravity, three Hu--Sawicki variants with $\left|f_{R0}\right| = 10^{-6},\,10^{-5},\,10^{-4}$ are included, which we denote as F6, F5, and F4, respectively (in order of increasing deviation from GR). 
For nDGP, two variants with $H_0 r_c = 5.0$ and $1.0$ are included, denoted as N5 and N1, respectively (again in order of increasing deviation from GR). 
Further details of the ELEPHANT simulations can be found in Ref.~\cite{cautun_santiago-harvard-edinburgh-durham_2018} and Ref.~\cite{alam_towards_2021}.

The haloes are identified by ROCKSTAR halo finder
\citep{behroozi_rockstar_2013}.
After that the galaxies are populated within halos utilizing the Halo Occupation Distribution method, following one of the 5-parameter HOD schemes suggested by
Ref.~\cite{zheng_galaxy_2007}. 
For GR, the HOD parameters adopted are from 
Ref.~\cite{manera_clustering_2013}
to match the galaxy clustering in the CMASS galaxy sample.
For modified gravity models, given that we have only one observable universe, the HOD parameters are tuned to approximate the same galaxy number density and projected galaxy two-point correlation function as observed in the case of GR.
The galaxy number density is around $n_g=3.2\times10^{-4}\, [h^{-1}\mathrm{Mpc}]^{-3}$ and the tuned $w_p(r_p)$ can be found in the Fig.~4 in Ref.~\cite{alam_towards_2021} for more detailed information.



\section{Marked Correlation Function}\label{MCF cal and result}

The screening mechanism enables modified gravity models to exhibit environmentally-dependent effects.
Although the projected two-point correlation functions of MG models have been tuned, we anticipate the emergence of deviations within the marked correlation function, given that MCF essentially represent an environment-weighted variant of the 2PCF.
A general MCF can be defined as
Ref.~\cite{white_marked_2016},
\begin{equation}
\calm(r)=\frac{1}{n(r)\bar{m}^2}\sum_{ij} m_im_j=\frac{1+W(r)}{{1+\xi(r)}}\ ,
\label{eqn:mcf}
\end{equation}
where the sum is over all pairs of a given separation $ r $, $n(r)$ is the number of such pairs, $\bar{m}$ is the mean mark for entire sample and $m_i$ is the mark for $i$th tracer. In the second equality, $\xi(r)$ denotes the standard two-point correlation function and $W(r)$ is the `weighted' correlation function.

In previous MCF studies, the mark definition was based on galaxy number density or the mass/Newtonian potential of their host halos. 
However, when the mark is defined by galaxy number density, which has been fine-tuned to match the observation, its overall discriminatory capability is degraded. While mass/potential based weighting methods require to estimate halo mass from observation. The mass estimation methods do exist and have seen ongoing improvements, accurately measuring the mass of galaxy host halos remains a complex challenge (e.g., Refs.~\citep{yang_galaxy_2007, yang_extended_2021}).

In this study, we propose a new type of the mark defined by the ``global'' gravitational potential, which distinguishes our approach from previous works. The reason why we define a ``global'' potential instead of a ``local'' one is that: the gravitational potential as mark in former works is calculated from the host halo, which neglects the effects of all other surrounded halos.
Ref.~\cite{shi_environmental_2017} found that the latter could significantly impact the overall gravitational potential, and thus should also be considered as well.
Note that in $\Lambda$CDM model this global gravitational potential can be directly computed from the galaxy distribution using Poisson equation, except the galaxy bias which will be normalized.

\subsection{Global gravitational potential estimation}

\begin{figure}
\centering
\includegraphics[width=9cm]{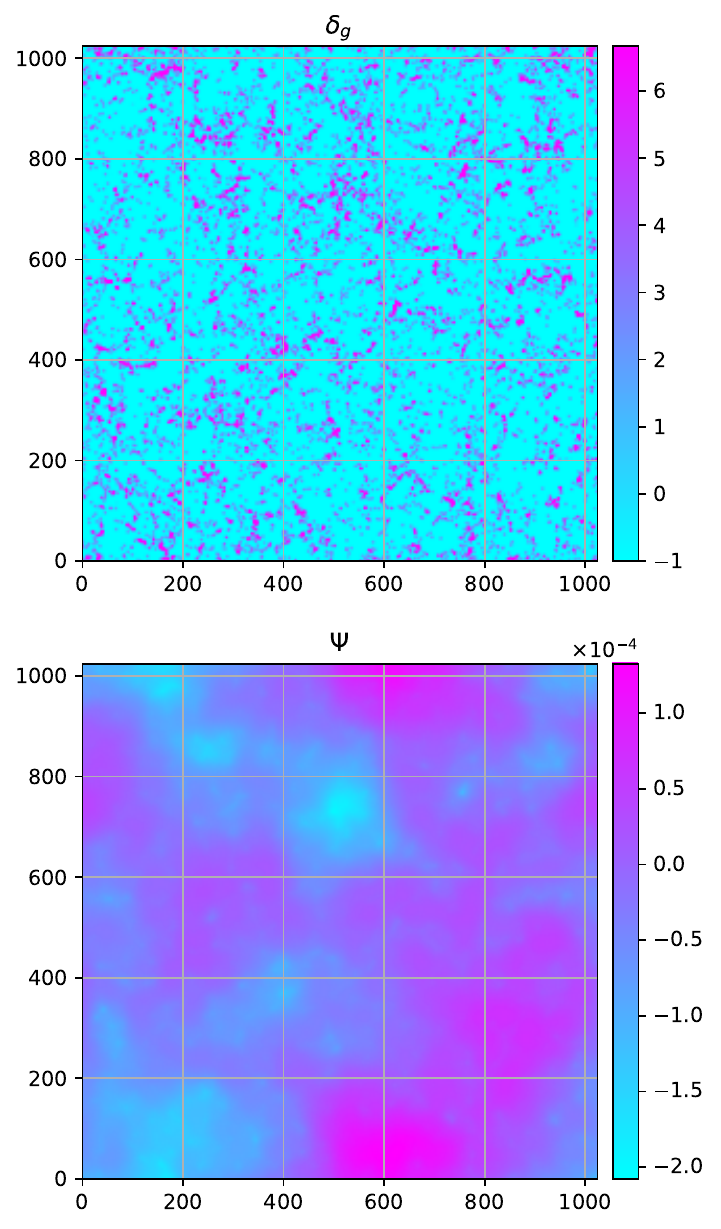}
\caption{2D slice of galaxy and potential distribution. The upper panel shows the slice of galaxy over-density field and the bottom panel shows the potential distribution calculated from Poisson equation. The results are from the first realization of GR model at $z=0.5$.}
\label{slice}
\end{figure}

\begin{figure}
\centering
\includegraphics[width=8cm]{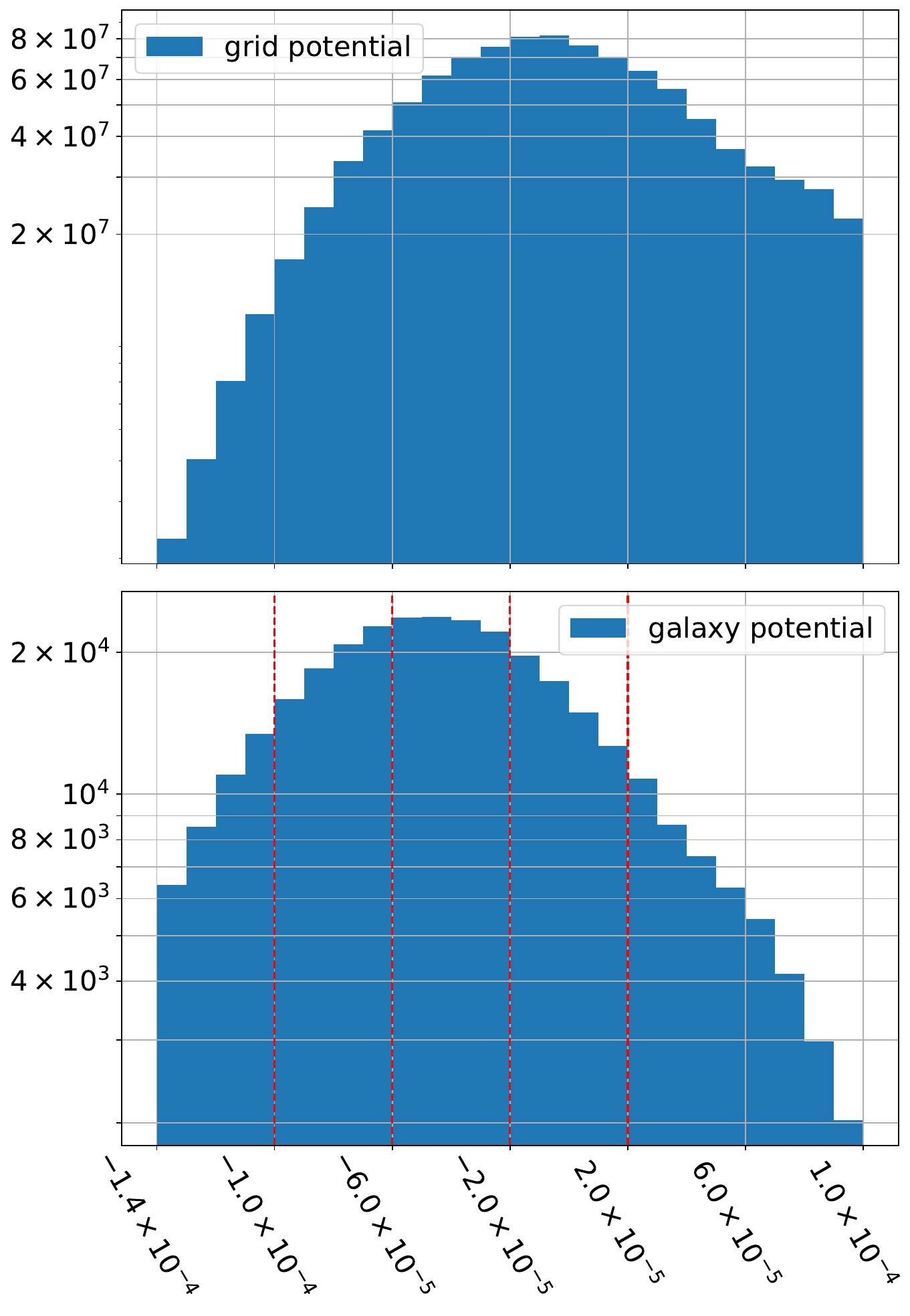}
\caption{Number of grids (upper panel) and galaxies (lower panel) with specified global gravitational potential values, which are estimated from the galaxy distribution.  The result is from realization 1 in GR model at $z=0.5$.
The negatively skewed peak in the lower panel is due to the fact that galaxies prefer to appear at high density and low potential region. The four dashed red lines in the bottom panel correspond to four potential thresholds used in step-function-type MCF in Section~\ref{StepFunc}.
}
\label{histo}
\end{figure}

The mark employed in this study is based on the gravitational potential field, which is obtained by solving the Poisson equation
\begin{equation}
-k^2\Psi=4\pi Ga^2\rho\delta\ ,
\end{equation}
where $k$ is the wavenumber in comoving units. 

In our methodology, we initially utilize the Cloud-in-Cell approach to generate the galaxy density field, denoted as $\delta_g$. This density is defined on a grid of $1024^3$ regular cells, each with a size of $1\ \mpch$. Subsequently, we employ the Fast Fourier Transform method to derive the final gravitational potential field.
In Fig.~\ref{slice}, we show a typical slice of galaxy density distribution and corresponding potential distribution from realization 1 of GR model. We find that high density region corresponds to negative potential well, which is in our expectation. 

Each galaxy is then assigned the gravitational potential of its nearest grid point, which serves as the basis for subsequent mark calculations.
In Fig.~\ref{histo}, we show the histogram of galaxy numbers as a function of assigned gravitational potential.
As a comparison, in the same figure we also show the number of cells with specific potential values.
The distribution of galaxies peaks at negative potential and shows strong asymmetry relative to zero point,
since the galaxies prefer living in high density and deep potential regions.

Note that a few approximations are adopted in the approach. First, we employ the galaxy density field as a substitute for the matter density field. 
At linear scale, the difference is the galaxy bias which is unknown in observation.
However, when we assign weights to galaxies according to their gravitational potentials, the key factor is the relative differences in potential rather than their absolute values.
Second, the same Poisson equation is applied to both GR and MG models, rather than using a modified version of the Poisson equation for each MG variants.
This is because we cannot precisely identify the gravity models of the Universe in advance. 

\section{Results}

\begin{figure}
\centering
\includegraphics[width=8cm]{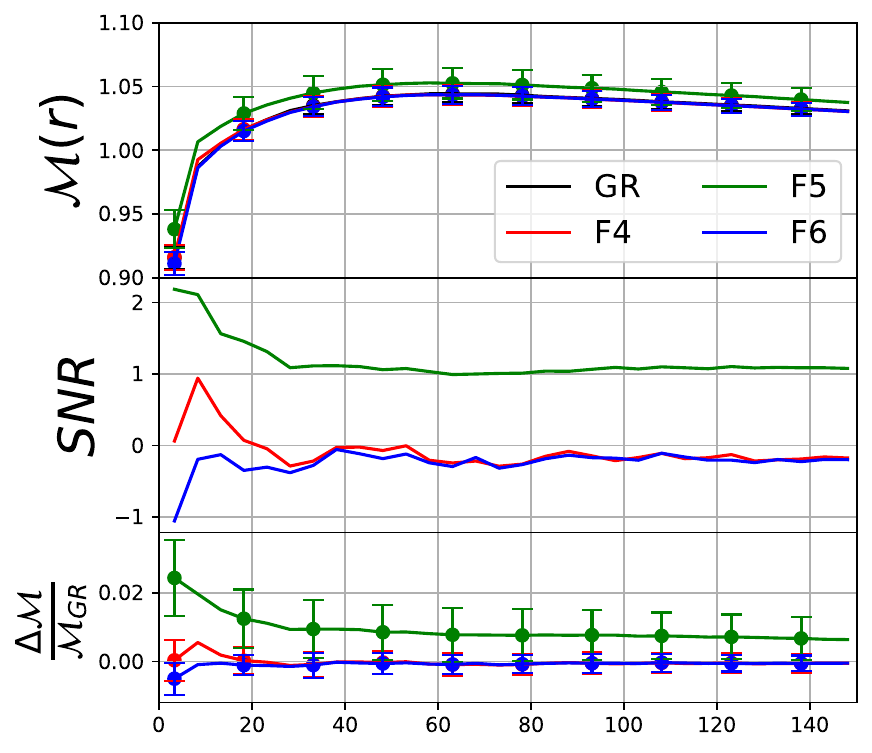}
\caption{Different forms of the MCF results for Sample 1 from the step-function-type marked galaxy correlation function. 
The upper panel shows $\mathcal{M}(r)$, the middle panel shows the signal-to-noise ratio (SNR), and the bottom panel shows the MCF fractional difference $\Delta M / M_{\mathrm{GR}}$. The error bars represent the standard error of mean over all 5 realizations.
Error bars are plotted for every three data points to enhance graph clarity.
}
\label{SNR_explain}
\end{figure}

We will investigate two types of mark which allow us to obtain the clustering information of galaxies in selected regions of interest.
We use \textit{nbodykit}\footnote{\url{http://nbodykit.readthedocs.io}} to calculate both the weighted and normal galaxy correlation functions in the nominator and the denominator of Eq.~\ref{eqn:mcf}, respectively.

In order to quantitatively assess the differences between MG and GR in the marked correlation functions, we first define the mean difference as
\begin{equation}
\overline{\Delta \mathcal{M}}(r) = \frac{1}{5} \sum_{i=1}^{5} \left\{ \mathcal{M}_{i,\mathrm{MG}}(r) - \mathcal{M}_{i,\mathrm{GR}}(r) \right\}\ ,
\end{equation}
where the index $i$ runs over the realizations. 
To estimate the statistical uncertainty of this mean difference, we estimate the standard error of the mean, given by
\begin{equation}
\sigma_{\mathrm{avg}}(r) = 
\sqrt{ \frac{1}{5 \times (5-1)} 
\sum_{i=1}^{5} \left\{ \Delta \mathcal{M}_i(r) - \overline{\Delta \mathcal{M}}(r) \right\}^2 }\ .
\label{eq:SEM}
\end{equation}
Finally, we define the SNR as
\begin{equation}
\mathrm{SNR}(r) = 
\frac{\overline{\Delta \mathcal{M}}(r)}{\sigma_{\mathrm{avg}}(r)}\ ,
\end{equation}
which provides a compact measure of the statistical significance of the difference between MG and GR MCFs.

In Fig.~\ref{SNR_explain}, we present three different forms of the MCF results: the marked correlation function
$\mathcal{M}(r)$, the signal-to-noise ratio $\mathrm{SNR}(r)$, and the MCF fractional difference $\Delta \mathcal{M} / \mathcal{M}_{\mathrm{GR}}$. 
Given that the uncertainties are highly correlated in correlation functions, 
the error bars are plotted for every three data points to enhance graph clarity for this plot and the similar plots hereafter.
We find that it is difficult to distinguish between the GR and MG models using only $\mathcal{M}(r)$. 
Therefore, to better highlight the differences, the subsequent results are presented in terms of the signal-to-noise ratio (middel panel) and the MCF fractional difference $\Delta \mathcal{M} / \mathcal{M}_{\mathrm{GR}}$ (lower panel). 


\subsection{Step Function Mark}
\label{StepFunc}
\begin{figure*}
	\centering
	\includegraphics[width=15cm]{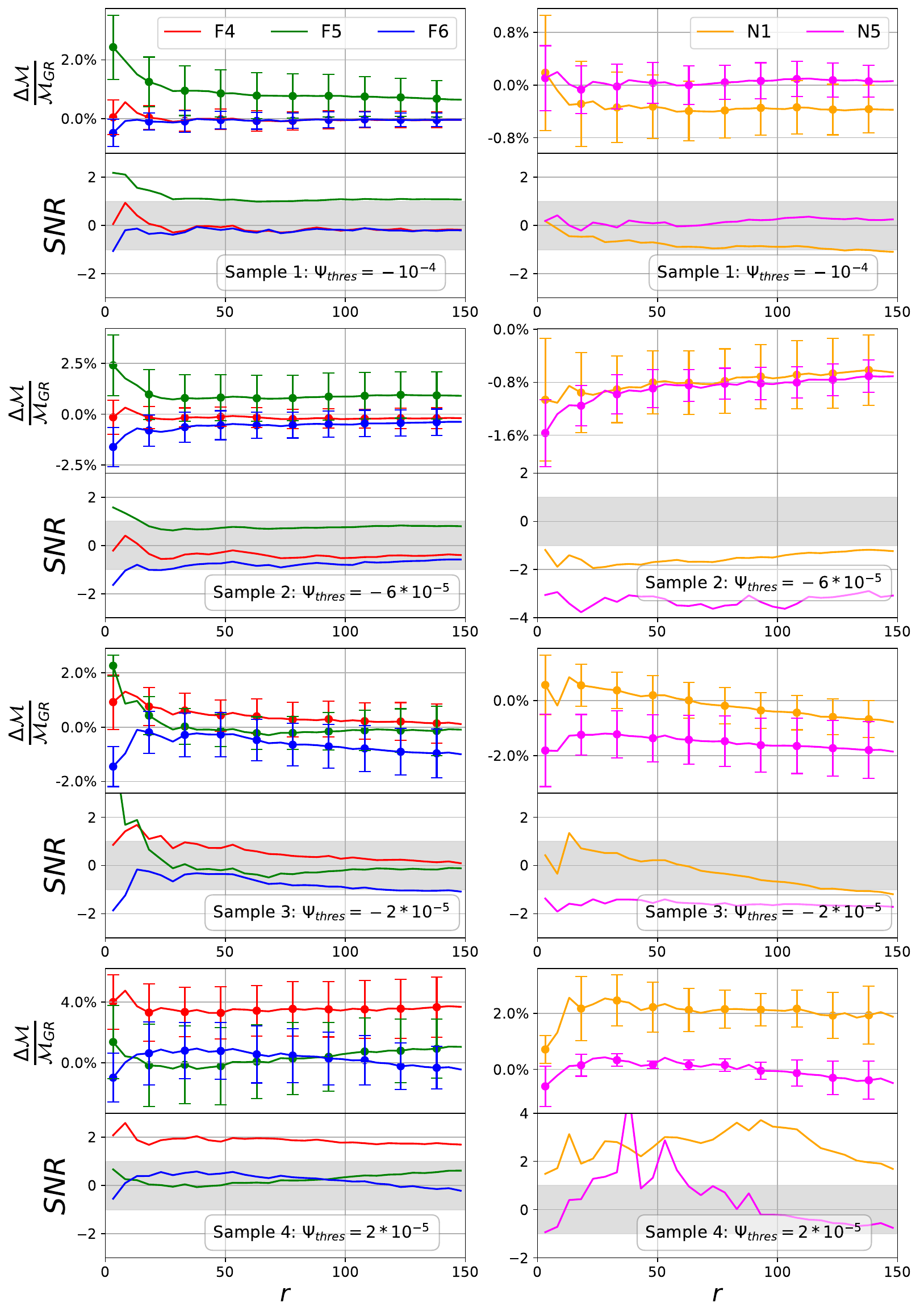}
	\caption{The signal-to-noise ratio (SNR) and the MCF fractional difference $\Delta \mathcal{M} / \mathcal{M}_{\mathrm{GR}}$ of step-function-type marked galaxy correlation function at $z=0.5$ for GR and three $f(R)$ variants/two nDGP variants (left/right panels). Plots from upper to bottom correspond to varying potential threshold $\Psi_\rmthres$. The error bars of MCF fractional difference represent the standard error of mean over all 5 realizations.
	}
	\label{All_step}
\end{figure*}


The first kind of mark is defined using step function,
\begin{equation}
m_i= \begin{cases}
1,\ \mathrm{if}\ \Psi_i \geq \Psi_{\rmthres}\ , \\
0,\ \mathrm{if}\ \Psi_i < \Psi_{\rmthres}\ .
\end{cases}
\end{equation}
where $\Psi_i$ stand for gravitational potential assigned to $i$th galaxy. This mark enables us to concentrate on galaxies situated in regions with a gravitational potential higher than a given threshold. From a more physical perspective, this weighting scheme can up-weight the unscreened areas.
We choose four thresholds with increasing value which correspond to four sample groups to investigate the dependence on the selection criteria.
Sample 1 has the largest number of galaxies living in various environments, while
Sample 4 has the lowest number of galaxies which mostly reside in the under-dense regions. We anticipate that there will be fewer discrepancies in Sample 1 between MG and GR. This expectation arises from the fact that Sample 1 only excludes a small portion of the entire galaxy sample set, aligning its $W(r)$ closely with the previous fine-tuned 2PCF, leading to the final MCF result of all gravity models very close to unity.
Conversely, we anticipate that the results in Sample 4 will exhibit more significant deviations between GR and MG. This is due to the galaxies in Sample 4 being significantly influenced by 
the additional fifth force from MG, potentially resulting in a distinct clustering pattern.
The results of step function type MCF are presented in Fig.~\ref{All_step}.

For F4 model, we observe increasing MCF fractional difference for increasing potential thresholds,
as well as the increasing SNR through Sample 1 to Sample 4 (red lines in top to bottom panels).
The deviation is negligible for Sample 1 and 2, and begins to be prominent in Sample 3 but only at small scales $r<30 \,\mpch$, and only at a level of $1\%$ for MCF fractional difference and $\sim1$ for SNR.
For Sample 4, the signal is very promising, with MCF fractional difference being $4\%$ and SNR being $\sim2$ over all scales.
For F5 results which are presented in green lines, we observe $1\%$ in MCF fractional difference and SNR $\sim1$ over Sample 1 and Sample 2 on all the scales we investigate. However, the signal disappear in Sample 3 and Sample 4.
For F6, only inconspicuous sign for modified gravity can be found.
The relatively large deviations at very small scale ($\sim-2\%$ in MCF fractional difference and SNR $\sim-1$) can be found for Sample 2, and at both small and large scale ($\sim-1\%$ in MCF fractional difference and SNR $~\sim-1$) for Sample 3.

Unexpectedly, results of F4 and F5 exhibit contrasting behaviors. In the case of F4, its results exhibit the largest deviation from GR in the highest potential threshold situation (Sample 4), whereas the smallest deviations from GR occur in the lowest potential threshold scenarios (Sample 1). This situation is entirely opposite for F5, with the largest deviations observed in situations with relatively lower potential thresholds (Sample 2) and the smallest deviations in cases with higher potential thresholds (Sample 1). When referring to the results of F6, we observe almost no detectable signal, as previously discussed. This could be simply attributed to the relatively weaker MG effects in F6 when compared to the other two $f(R)$ variants.

In Ref.~\citep{alam_towards_2021}, a similar non-monotonic dependence on the strength of \frtxt{} models was also noted in the galaxy bispectrum,
possibly attributed to the employed HOD model.
Additionally, this behavior was observed in both galaxy and halo MCF
\citep{hernandez-aguayo_marked_2018},
explained by the fact that the pronounced enhancement of gravity in F4 simulations induces increased halo mergers within dense regions, resulting in the enhanced formation of massive haloes. To maintain a constant number density $n_h$, more haloes from low-density regions must be incorporated into the halo catalogue, subsequently leading to reduced clustering.


For the N1 model, the behavior is complex. 
Sample 1 shows a $-0.4\%$ deviation across all scales, with the SNR remaining at a low level ($\mathrm{|SNR|}<1$). 
The MCF fractional difference increases to $-0.8\%$ in Sample 2, where the SNR rises to approximately $-2$. 
In Sample 3, the MCF fractional difference is nearly zero but exhibits clear scale dependence, showing a slightly positive signal at small scales and a slightly negative deviation at large scales. 
Similar to the F4 model, the most significant signal appears in Sample 4, where the MCF fractional difference reaches $2\%$ across all scales and the SNR approaches $2.5$. 
For the N5 model, both Sample 2 and Sample 3 show significant signals. 
In Sample 2, the MCF fractional difference ranges from $-1.6\%$ at small scales to $-0.8\%$ at large scales, with $\mathrm{|SNR|}>2.5$. 
In Sample 3, the MCF fractional difference lies between $-1.5\%$ and $-2.0\%$, with $\mathrm{|SNR|} \sim 2$.

\subsection{Rectangular Function Mark}

\begin{figure*}
	\centering
	\includegraphics[width=15cm]{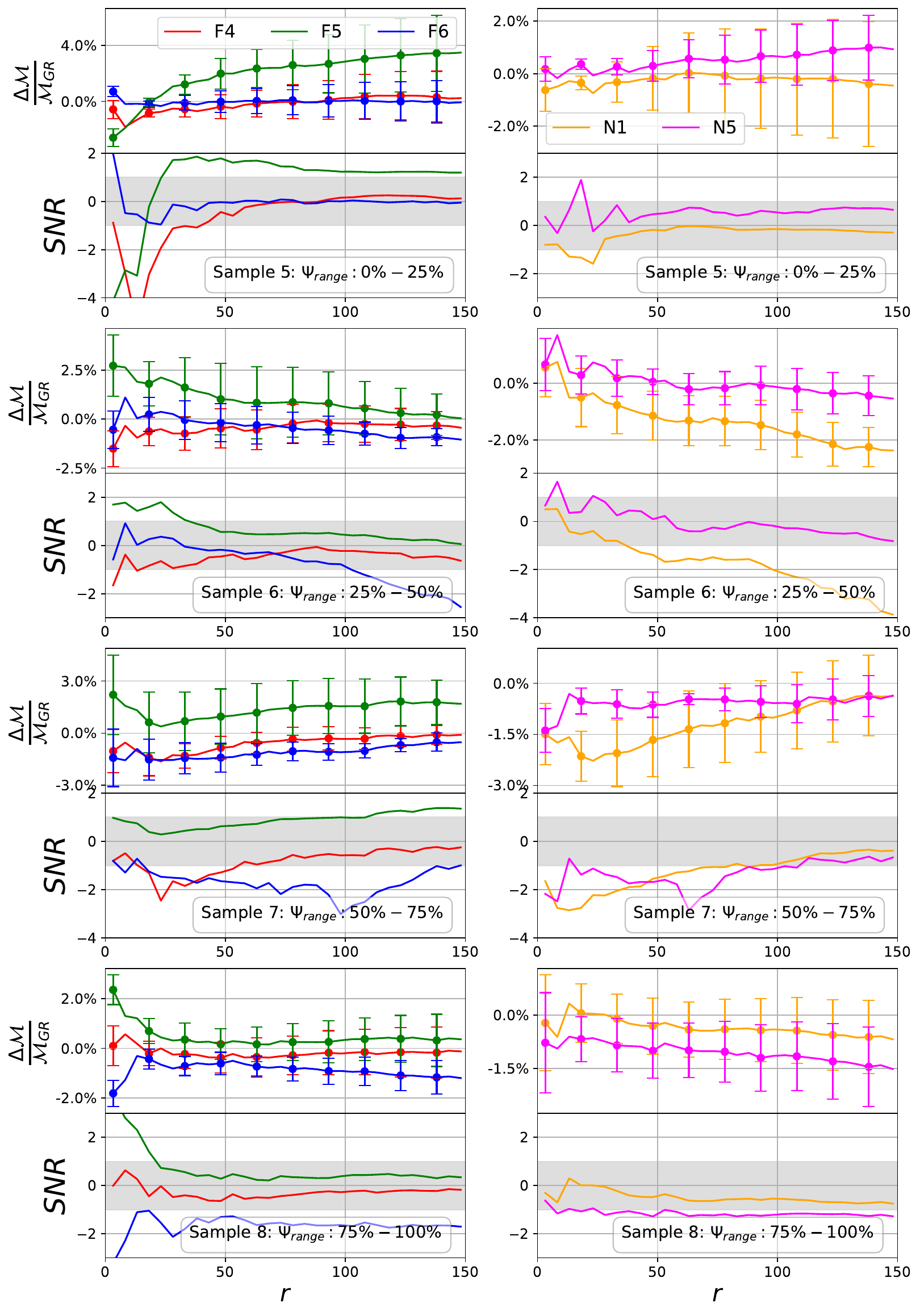}
	\caption{The signal-to-noise ratio (SNR) and the MCF fractional difference $\Delta \mathcal{M} / \mathcal{M}_{\mathrm{GR}}$ of rectangular-function-type marked correlation function of galaxy at $z=0.5$ for GR and three $f(R)$ variants/two nDGP variants (left/right panels). 
	Plots from upper to bottom correspond to setting the weighting in each quartile as one and zero otherwise. 
	The error bars of MCF fractional difference represent the standard error of mean over all 5 realizations.}
	\label{All_rect}
\end{figure*}


\begin{figure}
	\centering
	\includegraphics[width=8cm]{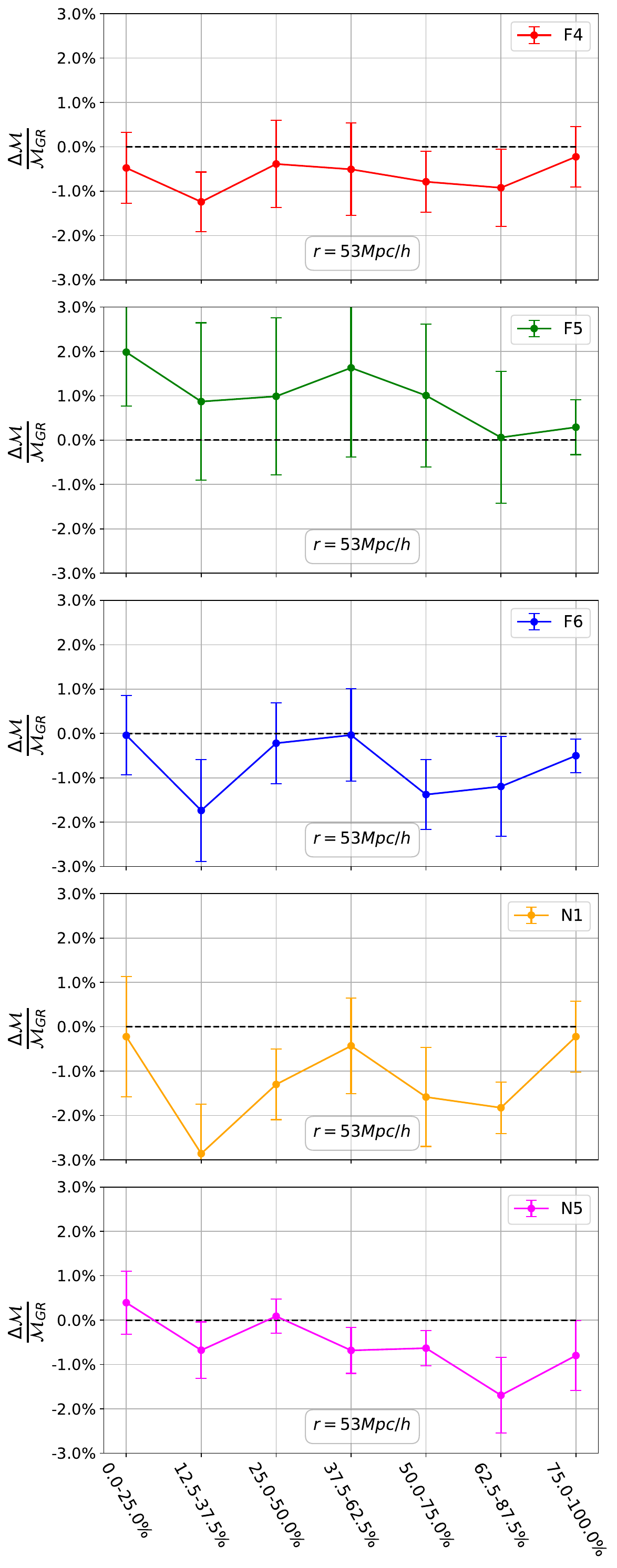}
	\caption{The rectangular-function-type marked correlation function at scale $r=53~\mpch$, with runing potential range $\Psi_{\rmrange}$ as x-axis. }
	\label{rcut}
\end{figure}

The second type of mark is based on rectangular function,
\begin{equation}
m_i= \begin{cases}
1, & \mathrm{if}\ \Psi_{\rmlower} \leq \Psi_i \leq \Psi_{\rmupper}\ ,\\
0, & \mathrm{otherwise}\ .
\end{cases}
\end{equation}
Here, $\Psi_{\rmupper}$ is the upper potential threshold and $\Psi_{\rmlower}$ is the lower one. This mark allows us to focus on the galaxies located in regions whose potential is in a specified range.
For simplicity and comparability, we choose the upper and lower limits to contain a fixed number of galaxies, rather than using two preassigned potential thresholds. The calculation of the MCF numerator covers a fixed 25\% of the total galaxies, with lower bound at 0\%, 25\%, 50\% and 75\%, respectively.
The results for the four quarters, also called Sample 5 to Sample 8, are displayed in Fig.~\ref{All_rect}.

From the top to the bottom panels on left side of Fig.~\ref{All_rect}, which correspond to different galaxy samples arranged in ascending ranges of gravitational potential, several significant deviations can be identified. 
For the F4 model, prominent signals are found only in Samples 5 and 7. 
In the case of F5, the absolute values of SNR for F5 are generally higher than those for F4 across most samples and scales, reaching $\mathrm{|SNR|}>2$ in Samples 5 and 8 at the smallest scale. 
For F6, we observe noticeable discrepancies in Samples 6, 7, and 8. Specifically, $\mathrm{|SNR|}$ reaching $\sim 2$ at the largest scale in Sample 6 and across all scales in Samples 7 and 8. 
This is in contrast to the step-function MCF result, in which the signal is diluted by including more galaxies living in relatively low potential region.
Despite these significant deviations, we are unable to identify a clear trend in the differences between the $f(R)$ variants and GR.

Next, we turn our attention to the results of the nDGP variants, as shown on right side of Fig.~\ref{All_rect}. 
For the N1 model, we observe higher absolute values of SNR in Sample 6 ($>2$ at large scales) and Sample 7 ($>2$ at small scales). 
However, in the remaining two samples, Samples 5 and 8, we do not find distinguishable discrepancies between GR and N1. 
For the N5 model, deviations from GR are evident in Sample 7, where the SNR reaches $\sim -2$ at certain scales. 
Moreover, in Sample 8, we observe a higher and more stable SNR for N5, representing an even larger deviation than that seen in N1.



For better illustration, we present the rectangular-function-type MCF in an alternative form. Fig.~\ref{rcut} displays the $\Delta \mathcal{M}/ \mathcal{M}_{\mathrm{GR}}$ at a fixed $r$ value, and as a function of the gravitational potential range. In the figure, we observe two distinct `valleys' within each model's curve. The bottoms of these two `valleys' are situated within the $12.5\%$-$37.5\%$ and $62.5\%$-$87.5\%$ potential ranges, respectively. This suggests that a relatively high potential range scheme ($62.5\%$-$87.5\%$) can offer more constraining power than the extreme high potential range scheme (75\%-100\%). This result, together with the result from 12.5-37.5\% scheme, contradicts our initial expectation: a higher potential range usually implies less effective screening. 

\subsection{Discussions}

\begin{figure}
	\centering
	\includegraphics[width=8cm]{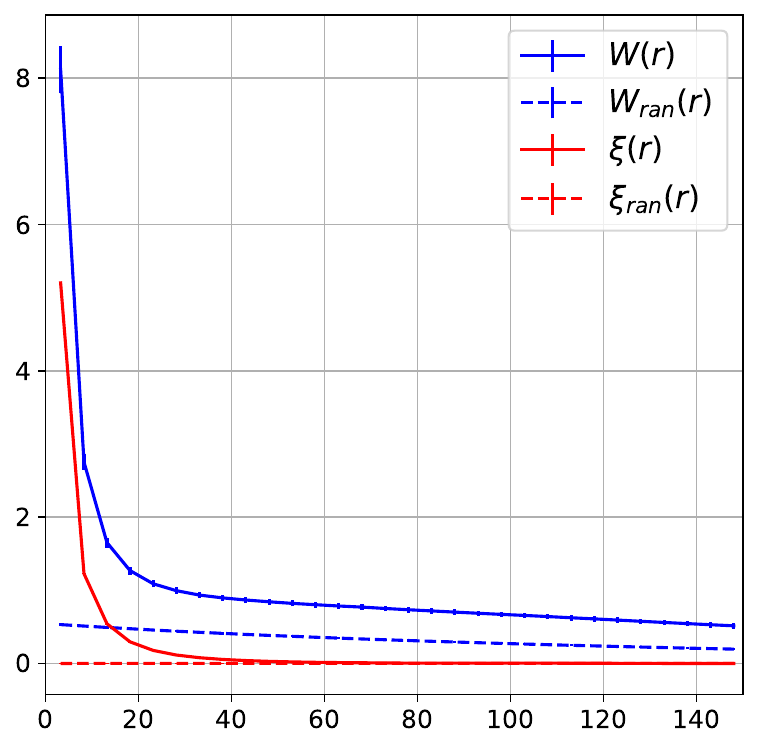}
	\caption{Comparison of weighted CF (WCF) and 2PCF of normal GR galaxy samples and random samples with subscript `ran'. The $W_{ran}(r)$ is calculated by using random galaxy catalogs but with the same global gravitational potential fields as those in the five GR samples. We choose $\Psi_{thres}=-2 \times 10^{-5}$ as threshold to calculate the step-function-type $W(r)$ and $W_{ran}(r)$.
	}
	\label{random}
\end{figure}

In Fig.~\ref{SNR_explain}, we find that $\mathcal{M}(r)$ does not approach unity on large scales for any of the models. This behavior also appears in nearly all subsequent results, although they are not shown here. This contrasts with the MCF reported in previous studies
\citep{alam_towards_2021,armijo_testing_2018,hernandez-aguayo_marked_2018}.
To delve deeper into this phenomenon, we respectively present the weighted correlation function $W(r)$ and the two-point correlation function $\xi (r)$ in Fig.~\ref{random}, which together constitute the final marked correlation function $\mathcal{M}(r)$. 
It becomes evident that the $1+\xi (r)$ approximates unity at large scale. Consequently, it is the denominator $W(r)$ that causes the final result $\mathcal{M}(r)$ to surpass unity at this scale.

Further, by applying global gravitational potential fields used in the GR galaxy samples to the random catalogs, 
we can separate the signal induced by the mark selection from the intrinsic galaxy clustering.
We observe that $W_{ran}(r)$ exceeds zero at large scales,
indicating that the large-scale correlation observed in the MCF above is partially attributed to the mark distribution. Additionally, $W(r)$ and $\xi (r)$
have similar trend at small scale, indicating that the small scale shape of $W(r)$ is mostly determined by the intrinsic galaxy clustering.
We observe $W(r)>W_{ran}(r)$ at large scale, as well as $W(r)>\xi(r)$ at small scale.
This boost is induced by the correlation between the galaxy distribution and mark distribution.
In the Poisson equation, the potential at a specific position is determined not only by local matter density but also by matter located far from that position. This leads to the coupling of the mark and the galaxy distribution.
On the contrary, marks defined by local properties, such as the local galaxy density or the host halo NFW potential in the aforementioned studies, lack this characteristic because they are inherently `local' and do not exhibit large-scale correlations.


This feature offers us the opportunity to utilize this MCF statistic to differentiate gravity models not only at small scale but also at large scale. 
In the previously mentioned works, using galaxy number density as mark only exhibits differences at scales $r<20 \ \mathrm{Mpc}/h$, while using host halo potential as mark only shows differences at scales $r<50 \ \mathrm{Mpc}/h$. 
Accurately modeling the results at intermediate to small scales is not an easy task, especially for the statistics beyond two point, such as MCF. Beyond that, there is also uncertainties introduced by the HOD modeling
\citep{armijo_new_2023}, which
predominantly affect the marked statistics on a small scale ($r_p<10h^{-1}$Mpc, see Fig.~7 in Ref.~\cite{armijo_new_2023}). 
However, deviations from GR in our results remain prominent at relatively larger scales. Consequently, we believe that these uncertainties may not significantly impact our final conclusion.

Rectangular-function-type marked correlation function in this work is closely related to density-split statistics mentioned in Ref.~\cite{alam_towards_2021}, since both methods isolate clustering signals from tracers within a selected range of an environmental variable rather than averaging over the full sample. In this sense, the rectangular mark is analogous to selecting a given percentile bin in density-split analyses. The difference is that density-split statistics still measure the clustering of explicitly separated subsamples, whereas the rectangular-type MCF remains within the marked-correlation framework and compares weighted clustering with the unweighted case. Moreover, density-split statistics are usually defined with the local density field, while our mark is based on the gravitational potential.


We also calculated the MCF for all the above samples in redshift space.
The results are quite similar in MCF monopoles, and thus we do not present the plots here. This is because that the global gravitational potential field is quite smooth, as shown in Fig.~\ref{slice}.
The marks obtained in real space and redshift space do not differ much.

\section{Conclusions}\label{Conclusion}
In this study, we have preliminarily examined the capability of the global gravitational potential weighted marked correlation function to constrain the modified gravity models using N-body simulations. We selected two representative modified gravity models: $f(R)$ gravity and the nDGP model. To match the observational results, the HOD parameters for these MG models were tuned to replicate the same galaxy number density and projected two-point correlation function. We employed the Poisson equation to generate the global gravitational potential and assigned this potential to each galaxy. We investigated two types of mark: the step-function-type and the rectangle or top-hat-function-type, and subsequently calculated the corresponding MCF for each gravity model.

For the step-function-type MCF, we find that generally a higher potential threshold enables MG models to exhibit a more pronounced deviation from GR. The deviations are more pronounced on smaller scales and less so on larger scales but still remain detectable. Regarding the rectangle-function-type MCF, the two distinct `valley' features observed in Fig.~\ref{rcut} suggest that focusing on regions with a specific gravitational potential range may assist in distinguishing between different gravity models. 


Considering that this statistic relies only on the galaxy density field, 
it can be readily merged into the data analysis pipeline for the ongoing and  upcoming spectroscopic galaxy surveys. 
Future observations will provide large-volume and high-density galaxy samples, enabling high-precision measurements 
of the marked correlation function across a wide range of scales.

In future work, several aspects can be further explored. 
First, a more systematic investigation of the optimal potential thresholds 
and mark definitions may improve the sensitivity of the statistic to 
different screening mechanisms in modified gravity models. 
Second, observational systematics such as 
survey geometry, varying number density across redshifts, and imaging systematics should be incorporated to assess the 
robustness of the method in realistic survey conditions. 
Finally, combining the global gravitational potential weighted marked 
correlation function with other large-scale structure probes, 
such as galaxy clustering, redshift-space distortions, and weak lensing, 
may further enhance the constraining power on modified gravity theories.

\section*{Acknowledgements}
We gratefully thank Baojiu Li for providing the simulation data.
This work was supported by the National Key R\&D Program of China (Grant Nos. 2023YFA1607800, 2023YFA1607801, 2023YFA1607802, 2025YFA1614103), the National Science Foundation of China (Grant Nos. 12595311, 12273020), the China Manned Space Project with No. CMS-CSST-2025-A04, the ``111'' Project of the Ministry of Education under grant No. B20019, and the sponsorship from Yangyang Development Fund.
This project is supported in part by Office of Science and Technology, Shanghai Municipal Government (grant Nos. 24DX1400100, ZJ2023-ZD-001).
This work made use of the Gravity Supercomputer at the Department of Astronomy, Shanghai Jiao Tong University, which 
enabled the large-scale experiments presented in this work.

\bibliographystyle{apsrev4-2}
\bibliography{cite}

\label{lastpage}

\end{document}